\documentclass[namedreferences]{SolarPhysics}
\usepackage[optionalrh,solaenum,natbib]{spr-sola-addons} % For Solar Physics 
\usepackage{graphicx}                    % For eps figures, newer & more powerfull
\usepackage{color}                       % For color text: \color command
\usepackage{url}                         % For breaking URLs easily trough lines
\usepackage[pdfborder={0 0 0 },urlcolor=blue,breaklinks]{hyperref} 
\ifx \doiurl  \undefined \def \doiurl#1{\href{http://dx.doi.org/#1}{\url{#1}}}\fi
\ifx \adsurl  \undefined \def \adsurl#1{\href{http://adsabs.harvard.edu/abs/#1}{\url{#1}}}\fi

\begin{document}

\begin{article}

\begin{opening}

\title{SPECTROSCOPY OF SOLAR PROMINENCES SIMULTANEOUSLY FROM SPACE AND GROUND}

%%%%%%%%%%%%%%%%%%%%%%%%%%%%%%%%%%%%%%%%%%%%%%%%%%%
%% Authors Names
%
\author{G.~\surname{Stellmacher}$^{1}$, E.~\surname{Wiehr}$^{2}$ and I.\,E. {Dammasch}$^{3}$}
%%%%%%%%%%%%%%%%%%%%%%%%%%%%%%%%%%%%%%%%%%%%%%%%%%%
%% Runningheads
% 
\runningauthor{G.\,Stellmacher, E\,Wiehr, I.\,E.\,Dammasch}
\runningtitle{Prominence emissions from space and ground}

%%%%%%%%%%%%%%%%%%%%%%%%%%%%%%%%%%%%%%%%%%%%%%%%%%%
%% Affiliations 
%
  \institute{$^{1}$ Institute d'Astrophysique, Paris, France,
               email: \href{mailto:stell@iap.fr}{stell@iap.fr} \\
             $^{2}$ Institut f\"ur Astrophysik, G\"ottingen, Germany,
               email: \href{mailto:ewiehr@astro.physik.uni-goettingen.de} 
                          {ewiehr@astro.physik.uni-goettingen.de}\\ 
             $^{3}$ Max-Planck-Institut f\"ur Aeronomie, Katlenburg-Lindau, 
               Germany, 
             }

%%%%%%%%%%%%%%%%%%%%%%%%%%%%%%%%%%%%%%%%%%%%%%%%%%%
%%% Abstract 
\begin{abstract}
We present a comprehensive set of spectral data from two quiescent solar 
prominences observed in parallel from space and ground: with the VTT, 
simultaneous two-dimensional imaging of H$\beta\,4862$\,\AA{} and 
Ca\,{\sc ii}\,8542\,\AA{} yields a constant ratio, indicating small spatial 
pressure variations over the prominence. With the Gregory, simultaneous 
spectra of Ca\,{\sc ii}\,8542\,\AA{} and He\,{\sc i}\,10830\,\AA{} were taken, 
their widths yielding $8000< T_{kin}< 9000$\,K and $v_{nth}<8$km/s. The 
intensity ratio of the helium triplet components gives an optical thickness of 
$\tau < 1.0$ for the fainter and $\tau \le 2.0$ for the brighter prominence. 
The $\tau_0$ values allow to deduce the source function for the central line 
intensities and thus the relative population of the helium $^3$S and $^3$P 
levels with a mean excitation temperature $T^{mean}_{ex} = 3750$\,K. 

With SUMER, we sequentially observed 6 spectral windows containing 
higher Lyman lines, 'cool' emission lines from neutrals and singly 
charged atoms, as well as 'hot' emission lines from ions like 
O\,{\sc iv}, O\,{\sc v},  N\,{\sc v}, S\,{\sc v} and S\,{\sc vi}. The 
EUV lines show pronounced maxima in the main prominence body as well 
as 'side-locations' where the 'hot' lines are enhanced with respect 
to the 'cool' lines. The line radiance of 'hot' lines blue-wards of 
the Lyman series limit ($\lambda<912$\,\AA{}) appear reduced in the 
main prominence body.This absorption is also visible in TRACE images 
of Fe\,{\sc ix/x}\,171\,\AA{} as fine dark structure which covers 
only parts of the main ('cool') prominence body. 

The Lyman lines show a smooth decrease of line widths and radiance with 
increasing upper level k = 5 through 19. For $5\le k \le8$ the level 
population follows a Boltzmann distribution with $T_{ex}>6\cdot 10^4$\,K; 
higher levels k$>8$ appear more and more overpopulated. The larger 
widths of the Lyman lines require high non-thermal broadening close 
to that of 'hot' EUV lines. In contrast, the He\,{\sc ii} emission is 
more related to the 'cool' lines.

\end{abstract}

%%%%%%%%%%%%%%%%%%%%%%%%%%%%%%%%%%%%%%%%%%%%%%%%%%%
%% Keywords
%

\end{opening}
%-------------------------------------------------

%%%%%%%%%%%%%%%%%%%%%%%%%%%%%%%%%%%%%%%%%%%%%%%%%%%
%% Sections
%
\section{Introduction}%\label{s:?} 

Solar prominences are spectacular manifestations of NLTE plasma emissions 
from atoms and ions coexisting in very different excitation states. Lines 
from neutral metals and their singly charged ions indicate a cool plasma 
component, whereas lines from higher charged ions such as 
O\,{\sc iv}, O\,{\sc v}, N\,{\sc v}, S\,{\sc vi} and S\,{\sc v} 
indicate a much hotter component. The 
spatial distribution of cold and hot plasma and its relation to the 
fine-structure ('threads', visible, e.g., in H$\alpha$ images) is an 
essential problem in prominence physics. The line shifts from bulk 
velocities show a good coherence over wide spatial scales largely 
independent of the line formation temperature. In contrast, the integrated 
line intensities show a good spatial correlation only for lines from atoms 
with similar ionization states.

Two opposed scenarios have been proposed to explain the distribution of 
cool and hot plasma in prominences: (i) cold threads, each surrounded 
by a hot transition layer where the observed EUV emissions originate or 
(ii) isothermal threads of different temperatures whose relative number 
may vary. The second view might be favored by the observations from UV 
rocket spectra of the eclipsed sun by Yang, Nicholls, and Morgan (1975), 
who found the prominence sizes to increase with temperature. Similarly, 
HRTS data from Wijk, Dere and Schmieder (1993) and SUMER data from 
de\,Boer, Stellmacher and Wiehr (1998) indicate that the lines from 
higher ionized atoms are preferentially visible in peripheral prominence 
parts.
                         
%________________________________________________________________
%
   \begin{figure}[h] 
   \hspace{0mm}
   \includegraphics[width=12cm]{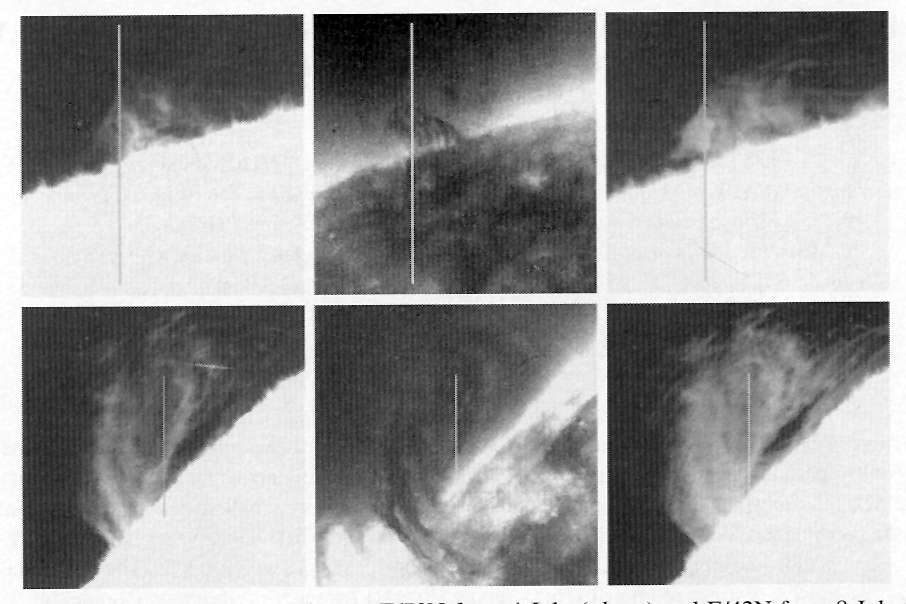}
   \caption{TRACE images of prominence E/70N from 4 July {\it (above)} 
and E/42N from 8 July 2000 {\it (lower panel)} in the 1600\,\AA{} 
continuum window {\it (left)}, in the coronal Fe\,{\sc ix/x}\,171\,\AA{} 
line showing absorption in the cool, dense prominence core {\it (middle)}, 
and in the Ly$\alpha$ line {\it (right panel)}; the 120'' long SUMER slit 
is indicated.}
%\label{Fig.1}
    \end{figure}
%________________________________________________________________
%

Observations in the visible spectral region mainly provide information 
about the cool parts, whereas EUV spectra particularly reflect emissions 
from hotter parts of a prominence. We present coordinated spectroscopic 
prominence observations from the ground, using at Tenerife the VTT and 
the GCT; in parallel we took EUV spectra from space with SUMER on board 
SOHO and two-dimensional images with TRACE.
We observed two prominences at the east limb, one 'polar crown type' 
at $70^o$\,N on 4 July 2000, and another at lower latitude $42^o$\,N on 
8 July 2000 (hereafter referred to as E/70N and E/42N).

\section{Ground-Based observations}

With the 
Vacuum Tower Telescope (VTT) on Tenerife two-dimensional images were 
taken simultaneously through narrow-band filters covering the integrated 
emissions of H$\beta$\,4862\,\AA{} and of Ca\,{\sc ii}\,8542\,\AA{}. The 
pressure-sensitive ratio of both lines is in agreement with our former 
observations (Stellmacher \& Wiehr 2000).Comparison with  model 
calculations by Gouttebroze and Heinzel (2002) for the optically thin 
case $\tau(H\beta)\le1.0$ (which is generally valid for quiescent 
prominences) shows that our data is compatible with a gas-pressure 
of $0.1<P_g<1.0$ dyn/cm$^2$.

In parallel we observed with the Gregory-Coude Telescope (GCT) on Tenerife 
spectra of the near infrared lines Ca\,{\sc ii}\,8542\,\AA{} and 
He\,{\sc i}\,10830\,\AA{} (hereafter referred to as Ca\,IR and He\,IR) 
simultaneously in the 5th and the 4th grating order of the spectrograph. 
The good spectral and spatial coherence of Ca\,IR and He\,IR, mainly 
visible in the similar shifts from bulk velocities (Fig.\,2), indicates 
a common emitting volume. 
                         
%________________________________________________________________
%
   \begin{figure}[h]   
   \hspace{7mm}
   \includegraphics[width=10.5cm]{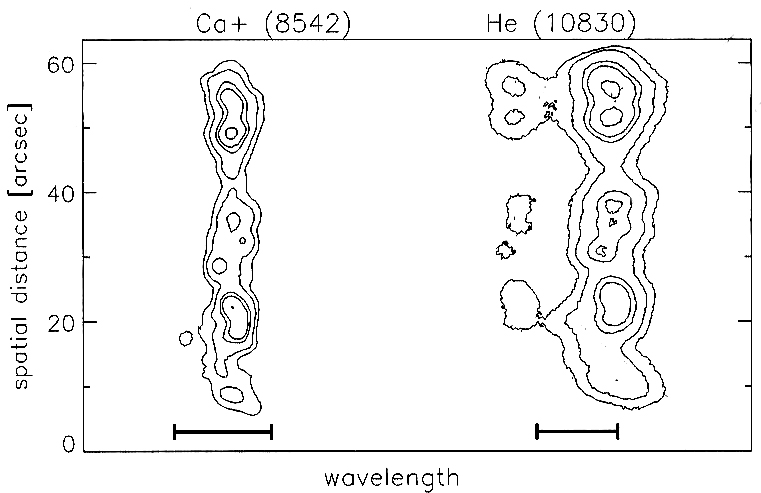}
   \caption{Iso-contours of the simultaneously observed emission lines 
Ca\,{\sc ii}\,8542\,\AA{} {\it (left side)} and He\,{\sc i}\,10830\,\AA{}
with the well separated blue triplet component in prominence E/70N; 
the bars give 1\,\AA{}; lambda increasing to the right}.
%\label{Fig.2}
    \end{figure}
%________________________________________________________________
%

\subsection{The Ca\,{\sc ii}\,8542 emission}

The widths of the Ca\,IR line can be considered as purely Doppler broadened:
due to the large atomic mass almost entirely by non-thermal (Maxwellian) 
velocities. For optically thin layers, the integrated emission ('line 
radiance') is proportional to the central line intensity: 
$E = \sqrt{\pi}\cdot\Delta\lambda_D\cdot I_0$. The lower limit of the 
$E$ versus $I_0$ relation yields minimum Doppler-widths 
$\Delta\lambda_D=120$\,m\AA{} and 130\,m\AA{} for prominences E/70N 
and E/42N, respectively. The observed emission relation (Fig.\,3) shows 
that each prominence is characterized by a well-defined relation; for 
the fainter prominence E/70N the slope is steeper than for the brighter 
one E/42N ('branching', cf., Engvold, 1978; Stellmacher and Wiehr, 1995).
                         
%________________________________________________________________
%
   \begin{figure}[ht] 
   \hspace{7mm}   
   \includegraphics[width=10.5cm]{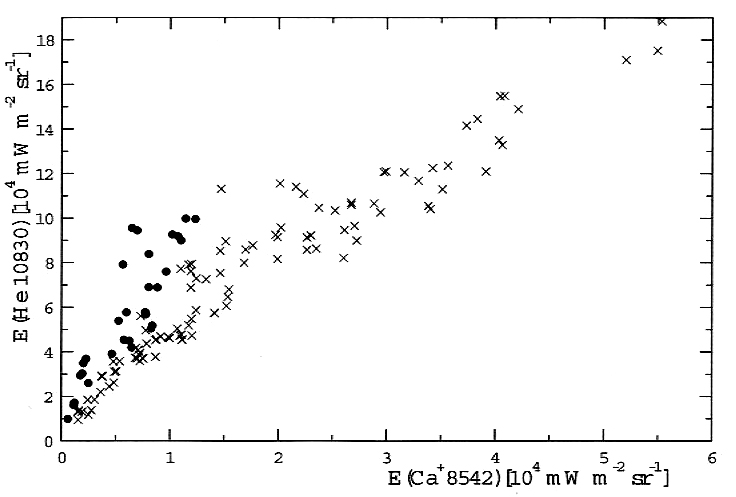}
   \caption{Observed emission relations of He\,{\sc i}\,10830\,\AA{} versus 
Ca\,{\sc ii}\,8542\,\AA{} for prominences E/70N {\it (dots)} and E/42N 
{\it (crosses)}.}.
%\label{Fig.3}
    \end{figure}
%________________________________________________________________
%

\subsection{The helium triplet}

The He\,{\sc i}\,10830\,\AA{} line triplet is composed of a blue component 
at 10829.08\,\AA{} ($^3P_o - ^3S_1$) and two red components at 
10830.25\,\AA{} ($^3P_1 -  ^3S_1$) and at 10830.34\,\AA{} ($^3P_2 - ^3S_1$).
They allow a simple evaluation of the total optical thickness, which is
described in more detail in the original version (Solar\,Phys. 217, 133)
of this paper. Since the Doppler width exceeds the 90\,\AA{} 
(fine-structure) separation of the two red components, these two lines 
form an unresolved broader line; the blue component, however, is well 
separated (cf., Fig.\,2). The ratio between the central line intensities 
of the combined (two) red and the faint (isolated) blue component, 
$I_0^{red}/I_0^{blue}$ is now calculated as a function of $\tau_0$, the 
optical thickness at the center of the strongest (red) triplet component; 
(the dependence on the width $\Delta\lambda_D$ is found to be negligible). 
                         
For very small $\tau_0$ a ratio of 8:1 is obtained, which corresponds 
to the relative values of the absorption oscillator strengths, $f_{ik}$. 
Increasing $\tau_0$ saturates the combined red components, while the blue 
component still grows, yielding a ratio of 3 : 1 for $\tau_0=2.0$. 
This calculated relation now allows us to deduce $\tau_0$ 
from the ratio of the line-center intensities $I_0^{red}/I_0^{blue}$, 
observed at a given prominence location. We can then relate the 
line radiance of the complete He\,IR triplet with the total optical 
thickness of the two (unresolved) red components and find that 
prominence E/70N has maximum values of $\tau^{red}=1.0$, whereas 
prominence E/42N reaches $\tau^{red}=2.0$.  

\subsection{Line widths}

Our profile calculations allow us to convert the observed line 
widths of the superposed two red components into true Doppler widths, 
$\Delta \lambda _D$, for which we find almost identical values as from 
the isolated (and optically thin) blue component. These widths show a 
close relation with the Ca\,IR widths (Fig.\,4), as to be expected 
from the good coherence of the He\,IR and Ca\,IR lines (Fig.\,2). 

%________________________________________________________________
%
   \begin{figure}[h] 
   \hspace{-2mm}   
   \includegraphics[width=12.2cm]{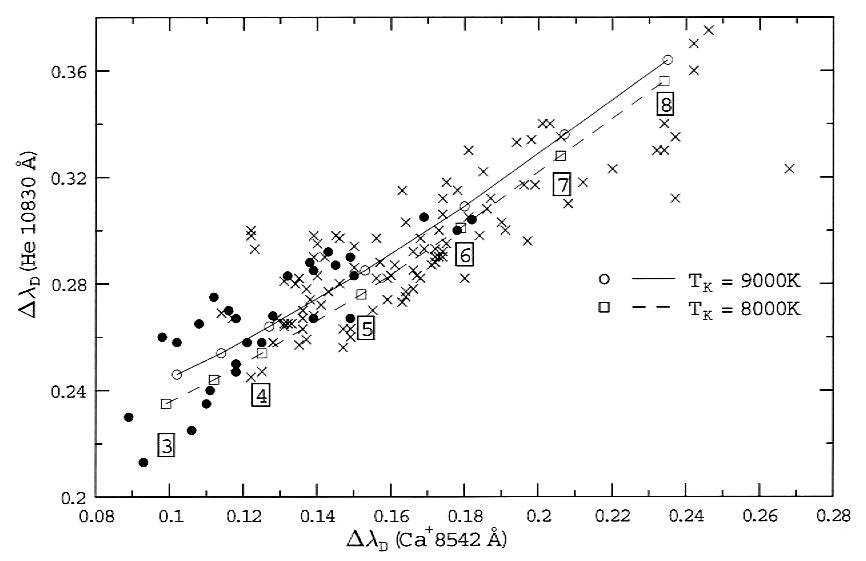}
   \caption{ Deduced dependence of the line emissions E(He\,{\sc i}\,10830\,\AA{}) 
on $\tau^{red}(= 1.56\tau_o)$ for the prominences E/70N {\it (dots)} and E/42N 
{\it (crosses)}.}.
%\label{Fig.4}
    \end{figure}
%________________________________________________________________
%

We superpose in Figure\,4 calculated Doppler widths, which show that the 
fainter prominence (E/70N) is characterized by $8000<T_{kin}<9500$\,K the 
brighter one (E/42N) being slightly cooler with $7500<T_{kin}<9000$\,K. 
This confirms earlier results by Stellmacher and Wiehr (1994, 1995) who 
found brighter prominences to be cooler (and less structured). 
\eject
The non-thermal velocities amount to $3<v_{nth}<6$\,km/s and $4<v_{nth}<8$\,km/s
for E/70N and E/42N, respectively. This agrees with the Doppler widths of
$\Delta \lambda _D$ (Ca\,IR) = 120\,m\AA{} and 130\,m\AA{}, respectively, 
obtained from the $E$ versus $I_0$ relation (Sec.\,2.1). Assuming purely 
non-thermal broadening, this corresponds to 4.2\,km/s and 4.5\,km/s for
E/70N and E/42N, respectively. Hence, the fainter prominence shows higher
$T_{kin}$ but smaller $v_{nth}$ than the brighter one.

\section{Space Observations}

In parallel with these ground-based observations, we took EUV spectra with
SUMER on board SOHO sequentially for the ranges 910-955, 1060-1100, 1225-1270,
1245-1295, 1295-1340, and 1540-1580 A. The total exposure time for each
spectrum was 120 s; the complete series of the six spectra was repeated 15
times over four hours. A slit of 1\,arcsec width and 120\,arcsec length was 
oriented along the solar north-south direction. The stray-light was taken 
from a sequence taken on 10 July where the SUMER slit did not cover a 
prominence; it was subtracted after careful alignment of the limb positions.

%________________________________________________________________
%
   \begin{figure}[h] 
   \hspace{-3mm}   
   \includegraphics[width=12.4cm]{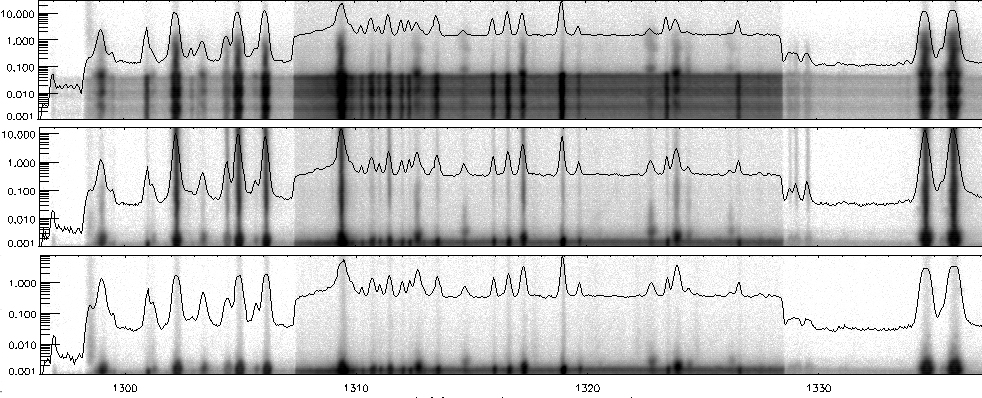}
   \caption{SUMER spectra in the range 1295-1340 \AA{} (first order) and 
648-670\AA{} (second order) for the prominences E/70N {\it (upper)}, 
E/42N {\it (middle)} and a region without prominence emission, 
taken for the background intensities {\it (bottom)}; the solar disk appears 
dark in the lower parts of the spectra.}.
%\label{Fig.5}
    \end{figure}
%________________________________________________________________
%

As an example of the SUMER spectra we show in Figure\,5 the range 
1295\,-\,1340\,\AA{} for both prominences in comparison to an off-limb 
region without prominence. It can be seen that ions of different formation 
temperature preferentially emit in different parts of the prominences. 

\subsection{The Lyman lines}

Among our large sample of EUV emissions, the Lyman lines are of particular
interest. Since damping does not play a significant role in prominences, the
emission lines can be fitted by Gaussian profiles in the wings. Figure\,6 
shows observed Lyman lines together with their respective fits. The lines 
are noticeably reversed for upper levels $5\le k\le7$ and still saturated for 
$8\le k\le9$. For these lines, the cores are kept as measured and only the 
wings are assumed Gaussian. For the higher Lyman members $10\le k\le 19$, 
the whole profiles are represented by Gaussians considering spectral line
superpositions. 
%________________________________________________________________
%
   \begin{figure}[ht] 
   \hspace{-2mm}   
   \includegraphics[width=12.3cm]{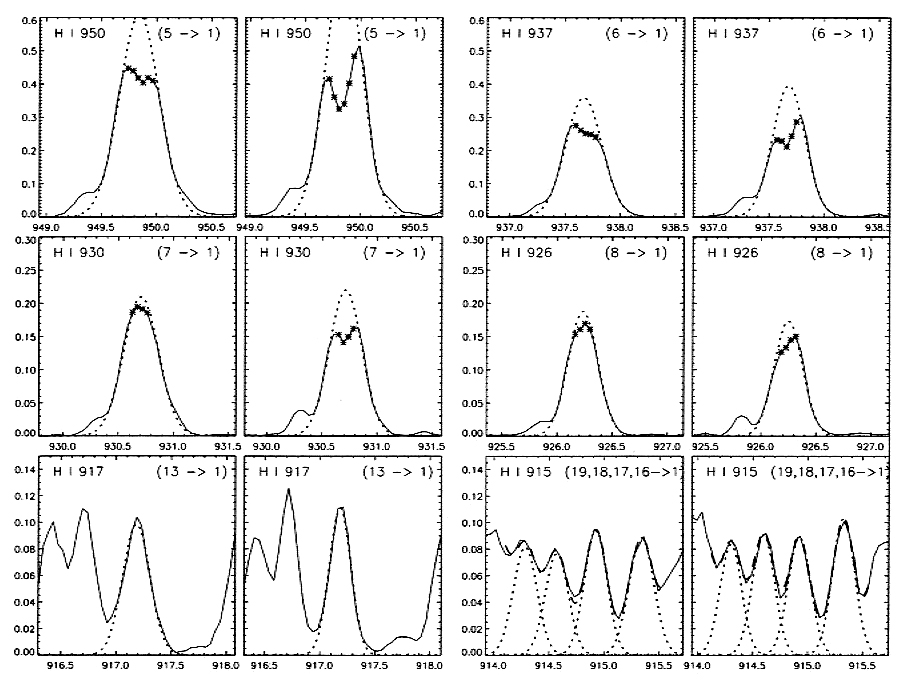}
   \caption{Observed Lyman line profiles with upper levels k = 5, 6, 7, 8, 
13, and 16...19; for prominence E/70N ({\it left}), prominence E/42N 
({\it right side}); Gaussian fits for $k<10$ only to the wings, the cores 
are taken from raw data {\it (asterisks)}; for $16\le k\le19$ the 
superposed Gaussians {\it (dashes)} deviate from the observed spectrum {\it 
(full line)} by typically 0.005...0.011.}.
%\label{Fig.6}
    \end{figure}
%________________________________________________________________
%

Our fits assure a highly accurate representation of the Lyman profiles and 
allow a reliable determination of their relative widths, 
$\Delta\lambda_e/\lambda$, and their line radiances (i.e., free from blends, 
mainly the He\,{\sc ii} lines). The thus obtained widths are plotted in 
Figure\,7 versus $log (\lambda f_{1,k})$ (being proportional to 
$log\,\tau_{1,k})$. Prominence E/42N exhibits narrower lines than prominence 
E/70N - in agreement with the 
ground-based results (Fig.\,4). Both curves are nearly parallel and decrease 
smoothly with increasing k and thus with decreasing optical thickness 
$\tau_{1,k}$. Even the smallest widths $\Delta\lambda_e/\lambda = 14 \cdot 
10^{-5}$ largely exceed $\Delta\lambda_D/\lambda = 4.3 \cdot 10^{-5}$ 
expected for Balmer lines with the $T_{kin}$ and $v_{nth}$ from our ground 
based spectra. The large widths of the Lyman lines can not arise from Stark 
broadening since we consider only upper levels k$<20$ (Hirayama, 1971). 
Instead, they indicate an origin from hotter layers of the prominence-corona 
transition region, PCTR.
\eject
The integrated line intensity ('radiance') versus $log (\lambda f_{1,k})$, 
shows a well defined relation ('curve-of-growth') indicating a common state 
of excitation. From the Lyman radiance, $E_{1,k}$, level populations along 
the line of sight, $log(n/g_k)\cdot d$, may be obtained following the relation 
for optically thin emissions:
$log(n_k/g_k) = log((E_{1,k}\lambda_{1,k})/(A_{1,k}g_{1,k}))$

%________________________________________________________________
%
   \begin{figure}[ht] 
   \hspace{-1mm}   
   \includegraphics[width=11.5cm]{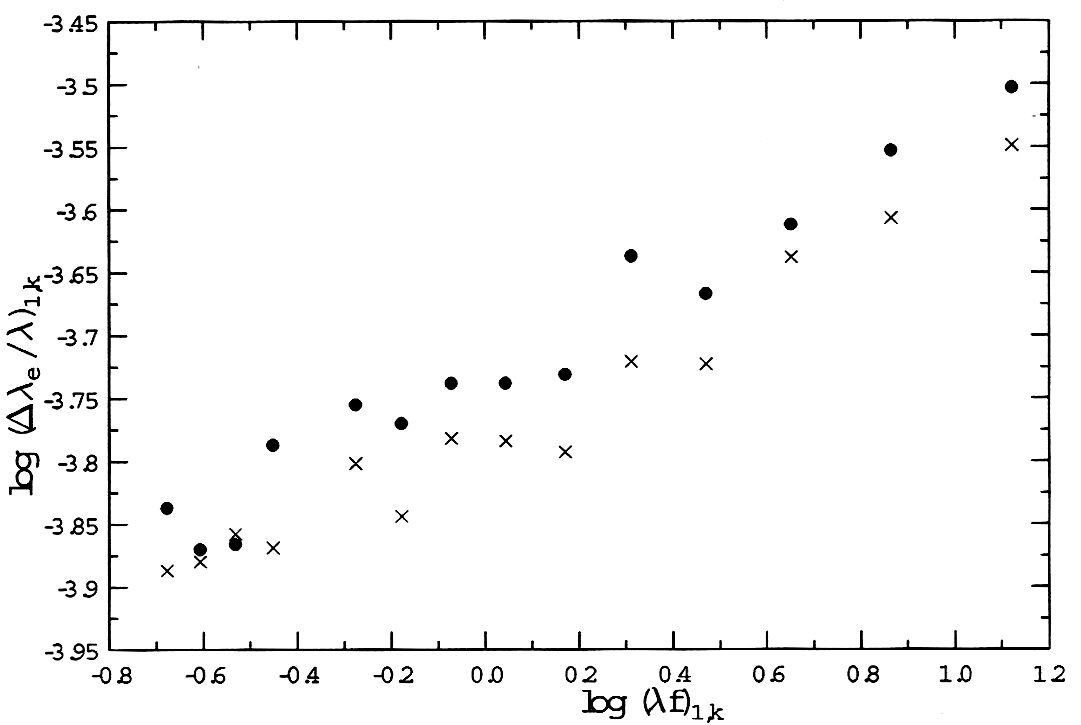}
   \caption{Averaged line widths at 1/e intensity as a function of the 
oscillator strength f for the Lyman members with upper levels $19 > k > 5$ 
{\it (from left to right: k = 15 omitted)} for prominences E/70N {\it (dots)} 
and E/42N {\it (crosses)}
.}.
%\label{Fig.7}
    \end{figure}
%________________________________________________________________
%

Plotting these values versus the corresponding energy distance from the
ionization limit $\Delta\chi=(\chi_{\infty}-\chi_k) = (13.6$ eV$-\chi_k)$ 
shows a smooth Boltzmann distribution for the strongly saturated lines (with 
pronounced central reversal) up to level $k=8$, which could be 
described by an equivalent excitation temperature of $T_{ex}>6\cdot 10^4$ K. 
The higher levels, $k>9$ with $\Delta\chi<0.2$\,eV (i.e., virtually the same) 
appear more and more overpopulated. Such a behavior can also be extracted 
from the data of Heinzel et al. (2001, cf., their Table 4).

\subsection{Spatial distribution of the EUV line intensities}

In agreement with our former study (deBoer, Stellmacher, and Wiehr, 1998), we
find in the present data a good spatial coherence of such emissions which
correspond to similar ionization states, respectively formation temperatures. 
For both prominences, the intensity scans along the slit in Figures\,8 and 9 
show local enhancements of the 'hotter' lines (S\,{\sc iv}, S\,{\sc v}, 
O\,{\sc v}; at 38'' in E/70N and at 28'' in E/42N) which are hardly seen in 
the 'cooler' lines (N\,{\sc i}, C\,{\sc i}, Fe\,{\sc ii} and Ni\,{\sc ii}).
\eject
Concerning the He\,{\sc ii} lines, we find that the spatial distribution of 
their emission is rather related to that of 'cool' lines as, e.g., 
Fe\,{\sc ii}\,1563\,\AA{}. Interestingly, the Lyman lines show a 
similar spatial variation (in Figures 8 and 9, Ly-10 is given as an 
example) and do not noticeably weaken at locations where lines from highly 
charged ions are strengthened. Since these regions can be considered to be 
preferentially composed of hot temperature plasma, their Lyman emissions 
will mainly reflect a recombination equilibrium. This might explain the 
small dependence of the intensity on k for higher Lyman members.

%________________________________________________________________
%
   \begin{figure}[ht] 
   \hspace{-4mm}   
   \includegraphics[width=12.4cm]{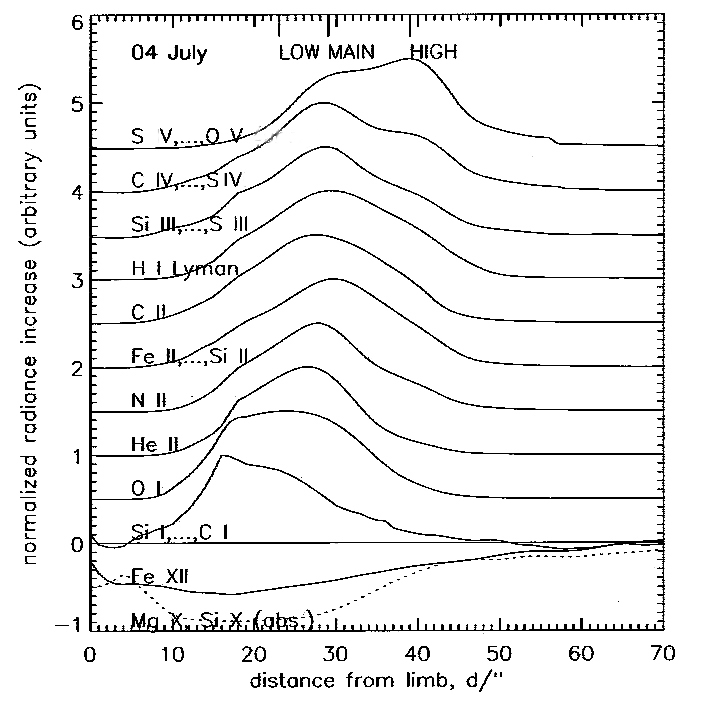}
   \caption{Spatial variation (along the SUMER slit) of the EUV emissions in  
prominence E/70N; counts integrated over the respective lines, arranged for 
increasing formation temperature {\it (upwards)}; negative values indicate 
absorbed coronal lines; the Lyman lines are represented by k = 10. The 7'' 
regions selected for spatial averaging are indicated at the {\it top} of 
the Figure.
}.
%\label{Fig.8}
    \end{figure}
%________________________________________________________________
%

%________________________________________________________________
%
   \begin{figure}[h] 
   \hspace{-2mm}   
   \includegraphics[width=11.5cm]{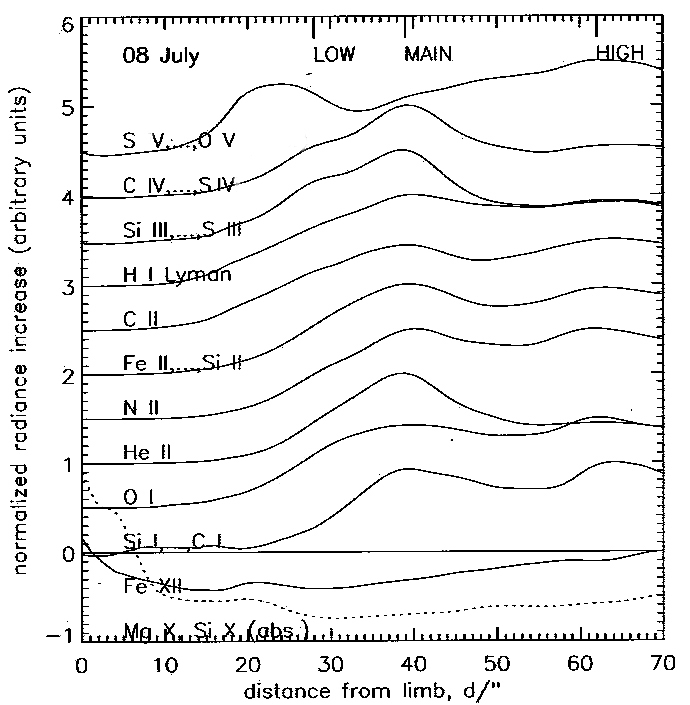}
   \caption{Same as Figure\,8 but for prominence E/42N
.}.
%\label{Fig.9}
    \end{figure}
%________________________________________________________________
%

\subsection{'Absorbed' EUV lines}

The purely coronal EUV lines (e.g., Fe\,{\sc xii}\,1242\,\AA{}; cf., lowest 
curves in Figures\,8 and 9) show an 'absorption' in the sense of reduced
radiance as compared to that from corresponding locations above the limb 
free from prominences. For coronal lines $\lambda<912$\,\AA{}, i.e. 
blue-ward of the Lyman series limit (e.g., Mg\,{\sc x}\,625\,\AA{}), an 
additional absorption occurs. Both cases can be seen in Figures\,8 and 9 at 
locations of maximum emission of the 'chromospheric' lines (e.g., Si\,{\sc i},
C\,{\sc i}). 
This relation of the EUV absorption to cool (dense) prominence matter is 
nicely seen comparing the Fe\,{\sc ix/x}\,171\,\AA{} image with the 
L$\alpha$ image from TRACE (Fig.\,1) and with the H$\alpha$ image from VTT.
In Figure 10 we give the profiles of the two 'coronal' lines 
Fe\,{\sc xii}\,1242\,\AA{} ($\lambda>912$\,\AA{}), and Mg\,{\sc x}\,625\,\AA{} 
($\lambda<912$\,\AA{}) in comparison with corresponding profiles from 
prominence-free locations at equal height above the solar limb. The additional 
absorption by the Lyman continuum is well visible for Mg\,{\sc x}\,625\,\AA{}. 

%________________________________________________________________
%
   \begin{figure}[h] 
   \hspace{0cm}   
   \includegraphics[width=12cm]{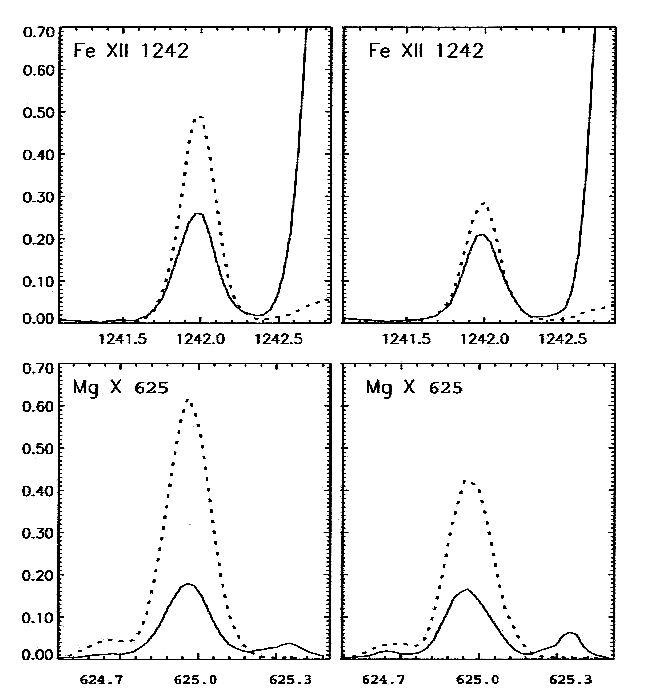}
   \caption{Emission profiles of two coronal lines in the prominence free 
background {\it (dashes)} and in the prominences {\it (full lines)} E/70N 
{\it (left panels)} and E/42N {\it (right panels)}, showing the absorption 
of Fe\,{\sc xii}\,1241.95\,\AA{} by the cool prominence material, and the 
additional absorption by the Lyman continuum of Mg\,{\sc x}\,624.95\,\AA{}
.}
%\label{Fig.10}
    \end{figure}
%________________________________________________________________
%

\subsection{Widths of the EUV lines}

The widths of even our narrowest observed Lyman lines 
$\Delta\lambda_e^{Ly}/\lambda\approx 14 \cdot10^{-5}$ largely exceed the 
Doppler widths $[\Delta\lambda_D/\lambda]_H= 4.3 \cdot 10^{-5}$ expected 
for the broadening parameters $T_{kin} = 8500$ K and $V_{nth} = 5$ km/s 
obtained from the ground-based spectra (Sec.\,2.3). The observed widths 
of the Lyman lines are closer to those of the 'hot' rather than of the 
'cool' (chromospheric) EUV lines in Figures\,11 and 12. This may be a 
further hint that the Lyman lines are emitted in prominence layers which 
are in contact with the hot corona (cf., above).

%________________________________________________________________
%
   \begin{figure} 
   \hspace{0mm}   
   \includegraphics[width=11.5cm]{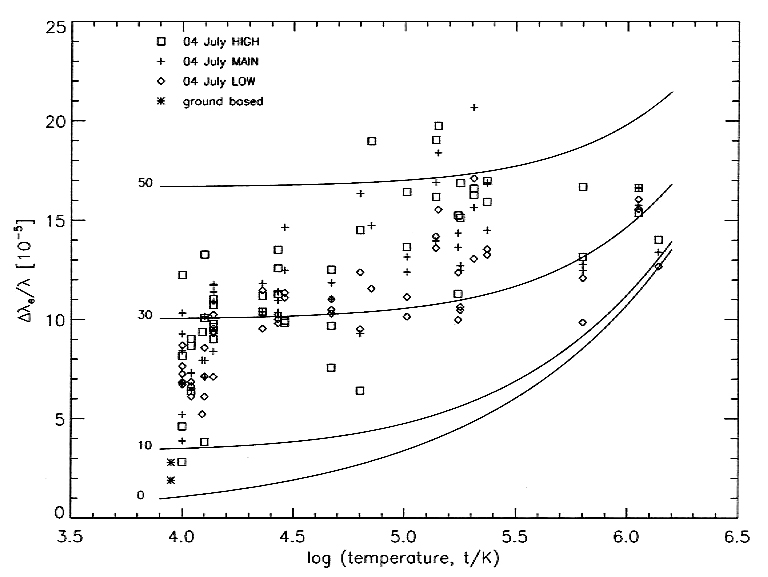}
   \caption{EUV line widths at 1/e intensity versus the formation temperature 
for the three spatial locations within prominence E/70N: the widths of the 
ground-based spectra of Ca\,{\sc ii}\,8542\,\AA{} and He\,{\sc i}\,10830\,\AA{} 
are also plotted (asterisks); for comparison Doppler widths calculated for the 
non-thermal velocities 0, 10, 30, 50 km/s and the (abscissa) temperatures with 
$\mu_{atom} = 16$ (oxygen) are shown
.}
%\label{Fig.11}
    \end{figure}
%________________________________________________________________

%________________________________________________________________
%
   \begin{figure} 
   \hspace{0mm}   
   \includegraphics[width=11.5cm]{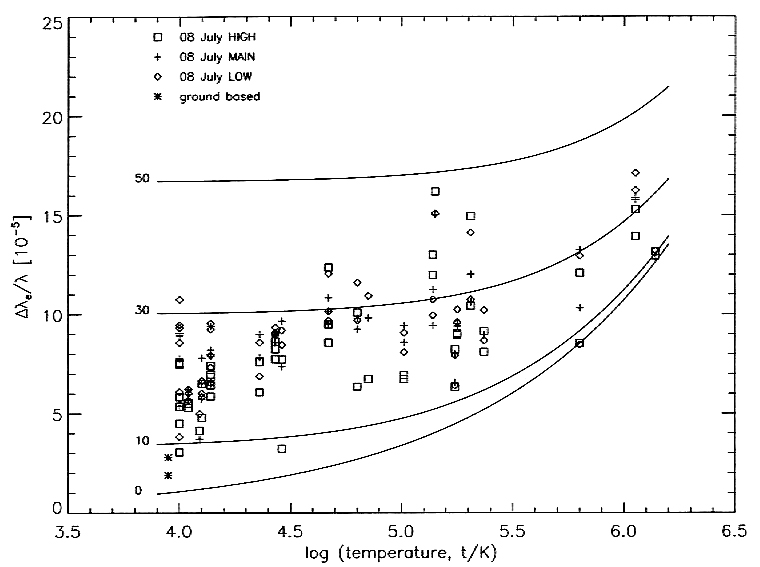}
   \caption{Same as Figure\,11 but for prominence E/42N
.}
%\label{Fig.12}
    \end{figure}
%________________________________________________________________

\subsubsection{Correction of narrow lines for the SUMER profile}

For the observed narrow EUV lines from neutrals and singly charged atoms 
(i.e., 'chromospheric lines'), the standard correction for the instrumental 
profile of detector\,A (cf., Wilhelm et al., 1997) yields reduced line widths 
$7\cdot 10^{-5}<\Delta\lambda_e/\lambda<12\cdot 10^{-5}$ for prominence E/70N 
and $5\cdot 10^{-5}<\Delta\lambda_e/\lambda<10\cdot 10^{-5}$ for prominence 
E/42N. If we apply the revised correction with broader instrumental profiles 
by Chae, Sch\"uhle, and Lemaire (1998), the resulting widths 
$5\cdot10^{-5}<\Delta\lambda_e/\lambda<9\cdot10^{-5}$ are still significantly 
broader than those obtained by Mariska, Doschek, and Feldman (1979) from 
observations with the NRL slit spectrograph on {\it Skylab} (at a spectral 
resolution of 60 m\AA{}). Interestingly, their reduced widths of lines from 
neutral and singly charged atoms of typically $\Delta\lambda_e/\lambda
\approx 3\cdot 10^{-5}$ are very near to those of our Ca\,IR line (entered in 
Figs.\,11 and 12).

We verified that the excess widths even in the maximum de-convoluted spectra 
from SUMER can not originate from our spatial averaging: the ground-based 
spectra do not yield sensibly wider profiles when equally averaged over 
7\,arcsec. The small dispersion of macro-shifts $< |5|$ km/s (cf., Fig.\,2) 
can not produce such a significant additional line broadening via spatial 
averaging. {\it We thus consider the observed narrow widths of EUV lines 
from neutral and singly charged atoms to be still affected by a underestimated 
instrumental profile of the SUMER spectrograph}, and conclude that we cannot 
draw definite conclusions from these widths of 'chromospheric' EUV lines.

\subsubsection{The broader 'hot' lines}

Almost insensitive to instrumental broadening are the wider lines from ions in
higher charged states (C\,{\sc iii}, C\,{\sc iv}, S\,{\sc iii} S\,{\sc iv} 
etc.), corresponding to formation temperatures $T_{form} > 10^{4.8}$ K (cf., 
Figs.\,11 and 12). Their widths smoothly increase with $T_{form}$. We consider 
this increase to be real, since Mariska, Doschek, and Feldman (1979) found 
comparable widths $\Delta\lambda_e/\lambda = 8 \cdot 10^{-5}$ and a similar 
increase with $T_{form}$, which they interpreted as a discontinuity of the 
non-thermal velocities between cool and hot emitting material. This increase 
of line widths is particularly pronounced in prominence E/70N (Fig.\,11) at 
location 'HIGH' where lines from higher-charged ions are enhanced (cf., 
Fig.\,8). At that location, the ground-based spectra show marked macro-shifts 
and a 'fuzzy' structure.

In order to give a reference for line broadening by temperature and by
non-thermal velocities, $v_{nth}$, we add in Figures\,11 and 12 curves of 
calculated Doppler widths as a function of the temperature (abscissa) for 
various values of $v_{nth}$, assuming the atomic mass of oxygen ($\mu = 16$). 
We see a marked difference between the two prominences: the emissions from 
higher charged ions give for prominence E/70N systematically larger $v_{nth}$ 
than for E/42N.

%________________________________________________________________
%
   \begin{figure} 
   \hspace{-4mm}   
   \includegraphics[width=12.4cm]{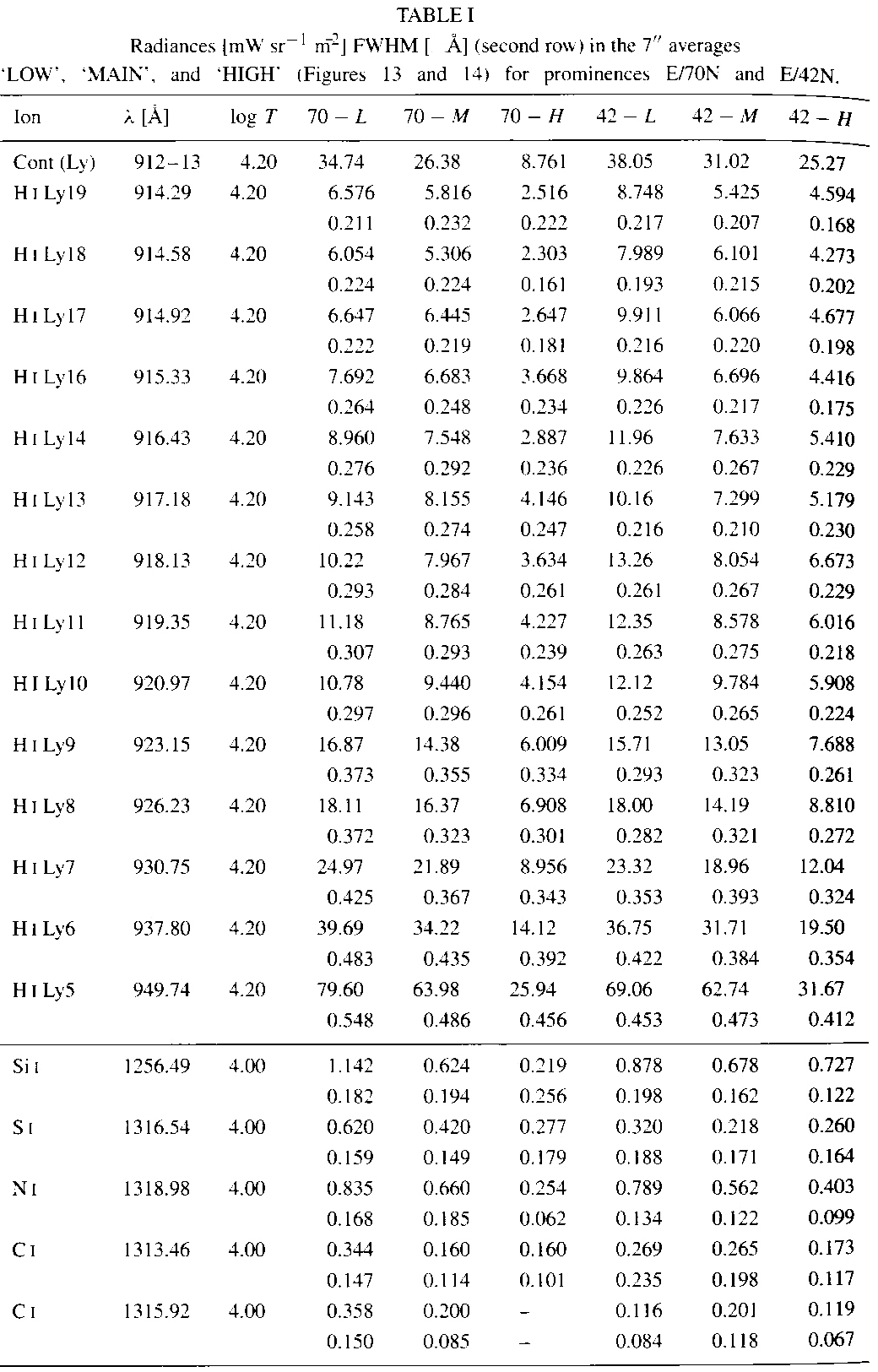}
%   \caption{.}
%\label{Fig.12}
    \end{figure}
%________________________________________________________________

%________________________________________________________________
%
   \begin{figure} 
   \hspace{-4mm}   
   \includegraphics[width=12.4cm]{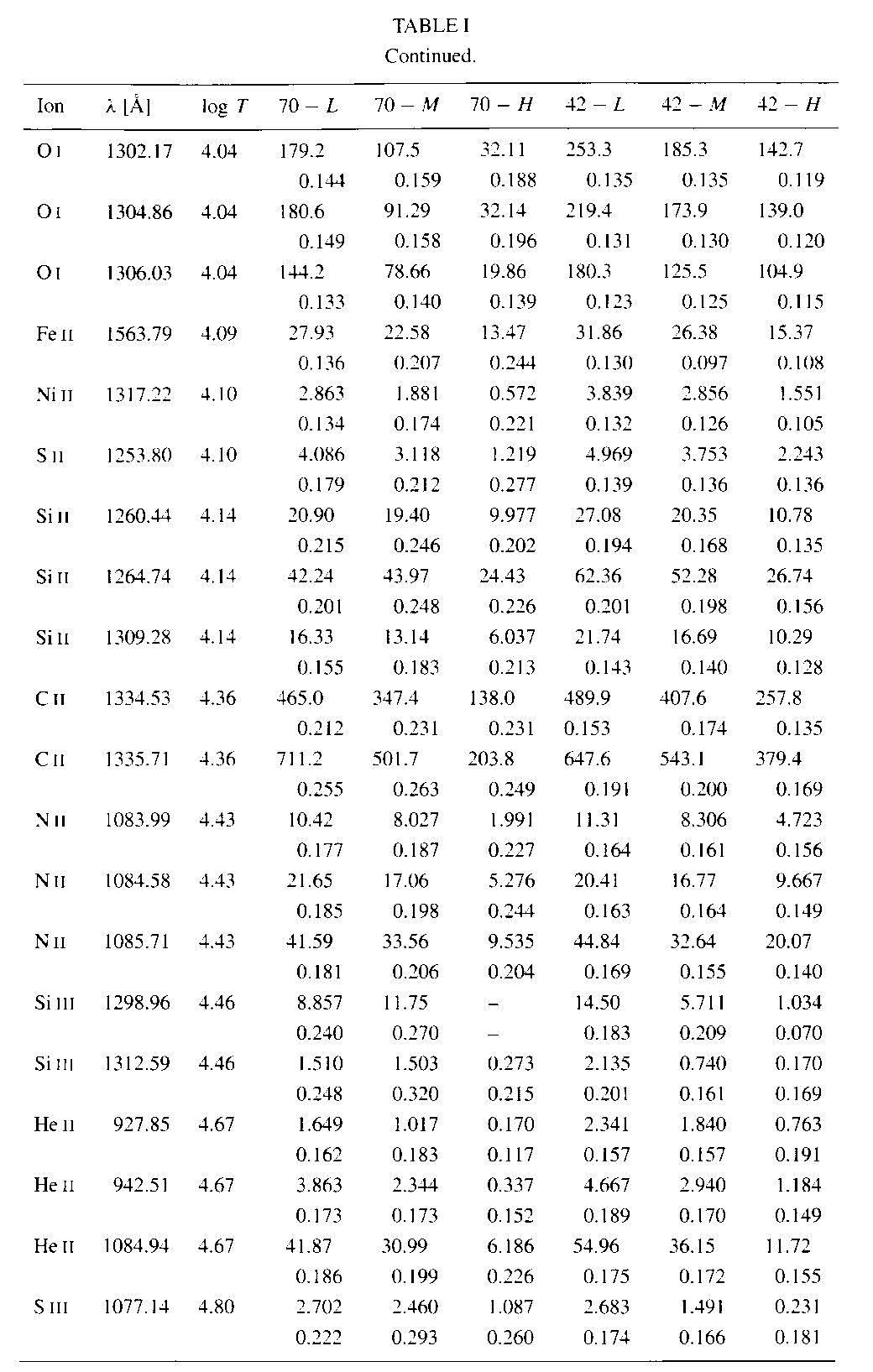}
%   \caption{.}
%\label{Fig.12}
    \end{figure}
%________________________________________________________________

%________________________________________________________________
%
   \begin{figure} 
   \hspace{-4mm}   
   \includegraphics[width=12.4cm]{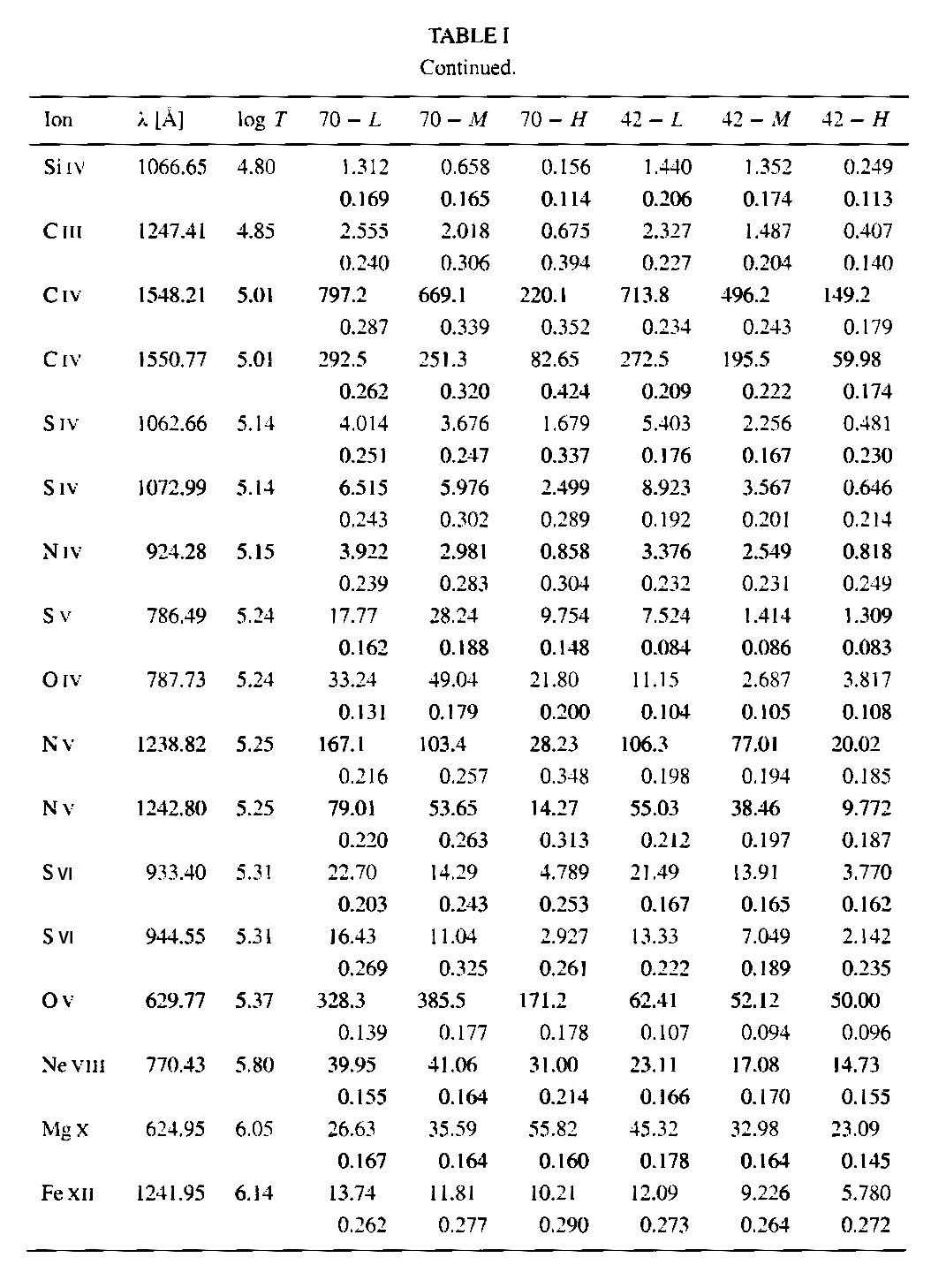}
%   \caption{.}
%\label{Fig.12}
    \end{figure}
%________________________________________________________________

\eject

\section{Conclusions}

The straightforward 'classic' determination of the plasma parameters kinetic
temperature, $T_{kin}$, and non-thermal (Maxwellian) velocity, $v_{nth}$, by 
comparison of line-widths from atoms with different mass and in same 
excitation state was not possible for the EUV emissions from atoms 
in lower excitation states (chromospheric lines), since the instrumental 
profile of the SUMER spectrograph is too broad and not precisely enough 
defined for those narrow prominence lines.

The He\,{\sc ii}\,1084\,\AA{} components (even though still affected by the 
SUMER instrumental profile) are so narrow that the He\,{\sc ii} formation 
has to be assumed close to that of the 'chromospheric' lines from neutrals 
and singly ionized atoms. This is also supported by the spatial similarity 
of the line radiances (Figs.\,8 and 9).

A comprehensive set of higher Lyman lines ($5\le k\le 19$) yields relative
widths $(\Delta\lambda_e/\lambda)^{Ly} > 14 \cdot 10^{-5}$ which smoothly 
increase with line strength (i.e., with optical thickness). Even their 
smallest widths largely exceed  $\Delta\lambda_e/\lambda = 4.3 \cdot 10^{-5}$ 
calculated for H\,{\sc i} lines with $T_{kin}= 8500$\,K and $v_{nth}=5$\,km/s 
from ground-based spectra. The well-defined emission relation of the Lyman lines 
(curve-of-growth) suggests a common state of excitation; their behavior 
significantly differs from that of the lines of the Balmer, Paschen, Bracket, 
and Pfund series, which typically yield $T_{kin}=6000$\,K a value, which would 
require for the observed widths of the Lyman lines non-thermal velocities 
$v_{nth}> 30$\,km/s 

The spatial resolution, though not matching that of the ground-based observations, 
allows us to distinguish regions with enhanced emissions of neutral and singly 
charged ions (mainly associated with 'cool' regions) from those of 'hotter' 
regions with enhanced emissions of lines from higher charged atoms
(such as O\,{\sc iv}, O\,{\sc v}, N\,{\sc v}, S\,{\sc v}, S\,{\sc vi}). This 
indicates that hotter locations exist separately from a cooler and denser main 
prominence region, as is qualitatively indicated from the two-dimensional 
images in Figure 1. All Lyman lines show a similar spatial variation, and do 
not noticeably weaken at locations where lines from highly charged ions are 
strengthened. This finding and the large widths of the Lyman lines suggest 
that their emissions will also originate from hot regions.

The 'discontinuity' between the widths of cool ('chromospheric') lines and
those from highly charged atoms (cf., Figs.\,11 and 12) indicates different
dynamical states of the emitting plasma, where not only the temperature but
also the non-thermal broadening is increased. This does principally not allow
us to distinguish between separate regions with 'hotter' and, respectively,
'cooler' emission or 'hot' transition regions surrounding 'cool' cores of 
each prominence thread. For a final decision between the various models, 
space and ground-based spectra should spatially resolve the prominence threads.

\eject

%%%%%%%%%%%%%%%%%%%%%%%%%%%%%%%%%%%%%%%%%%%%%%%%%%%%%%%%%%%%%%%%%%%%%%%%%%%
%% Acknowledgments
%
\begin{acks} 
We appreciate numerous discussions with W. Curdt, E. Marsch, and
K. Wilhelm, and support from the Max-Planck-Institut f\"ur Aeronomie, 
Katlenburg-Lindau; we thank an unknown referee for helpful suggestions. 
A. Garcia (IAP) and B. Bovelet (G\"ottingen) kindly performed some
of the graphics. The Gregory Coud\'e telescope on Tenerife is operated by the
Universit\"ats-Sternwarte, G\"ottingen (USG). and the Vacuum Tower Telescope by
the Kiepenheuer Institut f\"ur Sonnenphysik, Freiburg (KIS) at the Spanish
'Observatorio del Teide' of the Instituto de Astrof\'isica de Canarias. The
SUMER project is financially supported by the 'Deutsche Agentur f\"ur
Raumfahrt-Angelegenheiten' (DARA), the 'Centre National d'\'etudes Spaciales
(CNES), and the 'European Space Administration' (ESA).
\end{acks}

%%% %%%%%%%%%%%%%%%%%%%%%%%%%%%%%%%%%%%%%%%%%%%%%%%%%%%%%%%%%%%
%% Bibliography
%
% Using BibTeX
%
\bibliographystyle{spr-mp-sola}
% %\bibliographystyle{spr-mp-sola-cnd} %% Alternative style: no title, no concluding page
%\bibliography{helium}  
%
%\end{article} 
%\end{document}
% Without BibTeX 

\end{article} 
\end{document}